\documentclass[a4paper,fleqn,usenatbib]{mnras}
\usepackage{graphicx,url}
\usepackage{amsmath,amssymb}
\usepackage{hyperref}
\hypersetup{colorlinks,
	linkcolor=blue, 
	urlcolor=magenta, 
	anchorcolor=blue,
	citecolor=blue
}
\usepackage{bm}
\usepackage{xcolor}
\usepackage{mathrsfs}
\usepackage{natbib}
\usepackage{multicol}

\title[Determining the Cosmic Curvature with ML]{Machine Learning the Cosmic Curvature in a Model-independent Way}

\author[G.-J. Wang, X.-J. Ma and J.-Q. Xia]{
	Guo-Jian Wang,$^{1}$
	Xiao-Jiao Ma,$^{1}$
	and Jun-Qing Xia$^{1}$\thanks{E-mail: xiajq@bnu.edu.cn}\\
	$^{1}$Department of Astronomy, Beijing Normal University, Beijing 100875, China}

\date{Accepted: 2020 December 22; Revised: 2020 December 20; in original form 2020 April 30}

\begin{document}
\label{firstpage}
\pagerange{\pageref{firstpage}--\pageref{lastpage}}
\maketitle

\begin{abstract}
In this work, we achieve the determination of the cosmic curvature $\Omega_K$ in a cosmological model-independent way, by using the Hubble parameter measurements $H(z)$ and type Ia supernovae (SNe Ia). In our analysis, two nonlinear interpolating tools are used to reconstruct the Hubble parameter, one is the Artificial Neural Network (ANN) method, and the other is the Gaussian process (GP) method. We find that $\Omega_K$ based on the GP method can be greatly influenced by the prior of $H_0$, while the ANN method can overcome this. Therefore, the ANN method may have more advantages than GP in the measurement of the cosmic curvature. Based on the ANN method, we find a spatially open universe is preferred by the current $H(z)$ and SNe Ia data, and the difference between our result and the value inferred from Planck CMB is $1.6\sigma$. In order to test the reliability of the ANN method, and the potentiality of the future gravitational waves (GW) standard sirens in the measurement of the cosmic curvature, we constrain $\Omega_K$ using the simulated Hubble parameter and GW standard sirens in a model-independent way. We find that the ANN method is reliable and unbiased, and the error of $\Omega_K$ is $\sim0.186$ when 100 GW events with electromagnetic counterparts are detected, which is $\sim56\%$ smaller than that constrained from the Pantheon SNe Ia. Therefore, the data-driven method based on ANN has potential in the measurement of the cosmic curvature.
\end{abstract}


\section{Introduction}

In the field of cosmology, the cosmic curvature is a very important fundamental parameter. Many important problems such as the evolution of the universe and the property of the dark energy are closely related to the cosmic curvature. Based on theories of inflation, the radius of the curvature of the universe should be very large, thus, this implies that the cosmic curvature should be very small. Therefore, the detection of a significant deviation from $\Omega_K=0$ would have a profound impact on inflation models and fundamental physics. A lot of attention has been attracted to this issue \citep{Eisenstein:2005,Tegmark:2006,Zhao:2007,Wright:2007} and a flat universe is preferred by the latest Planck CMB experiment \citep{Planck2018:VI} with high precision: $\Omega_K=0.0007\pm0.0019$ (68\%,TT,TE,EE+lowE+lensing+BAO).

The cosmic curvature has been constrained with high precision, however, it should be emphasized that almost all of these estimations assume some specific models of dark energy, such as the equation of state of $w(z)$. Thus, these estimations are all model-dependent and indirect methods. It should be noted that due to the degeneracy between the spatial curvature $\Omega_K$ and the equation of state parameter $w(z)$ of dark energy, it is difficult to constrain these two parameters simultaneously. Thus, this hinders our understanding of the nature of dark energy. Therefore, cosmological model-independent estimations for the cosmic curvature will break the degeneracy and be helpful for studying the nature of dark energy.

Many works have been done to estimate cosmic curvature in a model-independent way. Based on the sum rule of distances along null geodesics of the Friedmann-Lema\^{\i}tre-Robertson-Walker (FLRW) metric, \citet{Bernstein:2006} proposed to constrain the cosmic curvature in a model-independent way. Then, it was used to test the validity of the FLRW metric \citep{Rasanen:2015}. However, the Union2.1 SNe Ia \citep{Suzuki:2012} and strong gravitational lensing systems selected from the Sloan Lens ACS Survey \citep{Bolton:2008} were used in their analysis. The large uncertainties of the gravitational lensing systems lead to a weak constraint on the cosmic curvature. Besides, the light curve fitting parameters of Union2.1 SNe Ia are determined by assuming the standard dark energy model with the equation of state being constant, thus, the constraint on the cosmic curvature is not completely model-independent. Another model-independent method was proposed to test the FLRW by combining the Hubble parameter measurements $H(z)$ and the transverse comoving distance $D_M(z)$ \citep{Clarkson:2007,Clarkson:2008}. In this method, the cosmic curvature can be estimated when the FLRW metric is valid:
\begin{equation}\label{equ:clarkson}
\Omega_{k}=\frac{\left[H(z)D_M'(z)\right]^{2}-c^{2}}{\left[H_{0}D_M(z)\right]^{2}} ~,
\end{equation}
where $c$ is the speed of light, $H_{0}$ is the Hubble constant, $D_M(z)=(1+z)D_A(z)=D_L(z)/(1+z)$ is the  transverse comoving distance \citep{Hogg:1999}, and $D^\prime_M$ is the derivative of $D_M$ with respect to redshift $z$. In the literature, this method has been widely used to test the FLRW metric or to estimate the cosmic curvature \citep{Cai:2016,Li:2014,LiShiYu:2019,Mortsell:2011,Rana:2016,Sapone:2014,Shafieloo:2010,Yahya:2014,Zheng:2019}. However, the derivative of the transverse comoving distance to redshift, $D^\prime_M(z)$, will introduce a large uncertainty.

To avoid shortcomings of the methods above, \citet{Yu:2016} proposed to measure the cosmic curvature by combining the transverse comoving distance $D_M$ and the proper distance $d_P$. In their analysis, the Hubble parameter measurements $H(z)$ and the angular diameter distance $d_A$ of baryon acoustic oscillation (BAO) is adopted. However, their analysis is not completely model-independent because $d_A$ and some of the Hubble parameter measurements are obtained from the BAO observations that dependents on the assumed fiducial cosmological model. Furthermore, using $H(z)$ and SNe Ia, \citet{Li:2016} and \citet{Wei:2017} proposed to measure the cosmic curvature in a model-independent way, and this method was applied to study the cosmic curvature and opacity \citep{Wang:2017}. In addition, \citet{Wei:2018} proposed to test the cosmic curvature by using the Hubble parameter and future gravitational waves (GW), \citet{Liao:2019} proposed to constrain the cosmic curvature with the lensing time delays and GW, \citet{Collett:2019} proposed to constrain on the cosmic curvature with the lensing time delays and SNe Ia, \citet{Wei:2019} proposed to constrain on the cosmic curvature using quasars and cosmic chronometers, and even with the strong lensing systems \citep{QiJZ:2019a,QiJZ:2019c,WangB:2020}.

In most of these papers, the Gaussian process (GP) \citep{Seikel:2012a}, a non-parametric smoothing method for reconstructing functions from data is used to construct a function of $H(z)$. GP is widely used in many recent works \citep{Bilicki:2012,Busti:2014,Cai:2016,Seikel:2012a,Seikel:2012b,Shafieloo:2012,Seikel:2013,WangDeng:2019,Wei:2017,Yahya:2014,Yang:2015,Yu:2016,Zhang:2016}. However, \citet{Zhou:2019} propose that the GP should be used with caution for $H(z)$ reconstruction. Moreover, both \citet{Wei:2017} and \citet{Wang:2017} find that GP is sensitive to the prior of the Hubble constant $H_0$, and the results are greatly influenced by the setting of $H_0$. This may imply that GP is unreliable when reconstructing $H(z)$.

Recently, \citet{Wanggj:2020} present a new non-parametric approach to reconstruct functions from data with Artificial Neural Network (ANN) and a public code {\sc R{\rm e}FANN}\footnote{\url{https://github.com/Guo-Jian-Wang/refann}} is developed. The ANN method has no assumption to the data and is a completely data-driven approach. In their analysis, the Hubble parameter can be reconstructed accurately without bias, and the function of $H(z)$ reconstructed by ANN can be used to estimation cosmological parameters. More importantly, the reconstructed function of $H(z)$ is not sensitive to the setting of the Hubble constant. Thus, they proposed that the data-driven method based on ANN will be a promising method in the function reconstruction.

In this work, we constrain the cosmic curvature in a model-independent using the Hubble parameter measurements $H(z)$ and SNe Ia, without assuming any fiducial cosmology. In our analysis, two nonlinear interpolating tools, ANN and GP, are used to reconstruct the Hubble parameter. We test the performance of ANN and GP in the measurement of the cosmic curvature. In addition, we also test the capability of the future gravitational waves (GW) standard sirens in the measurement of the cosmic curvature.

This paper is organized as follows: In section \ref{sec:methodology}, we reconstruct functions of the observational Hubble parameter using the ANN and GP methods. Section \ref{sec:constraint_omk} present the application of the reconstructed function of $H(z)$ in the constraint on the cosmic curvature. In section \ref{sec:discussion}, a discussion about the cosmic curvature is presented. Finally, a conclusion is shown in section \ref{sec:conclusion}.

\section{Reconstruction of $H(z)$}\label{sec:methodology}

In this section, we reconstruct functions of $H(z)$ by using the ANN method and the GP method. We will first introduce the Hubble parameter measurements and then achieve the reconstruction of $H(z)$.

\begin{figure*}
	\centering
	\includegraphics[width=0.32\textwidth]{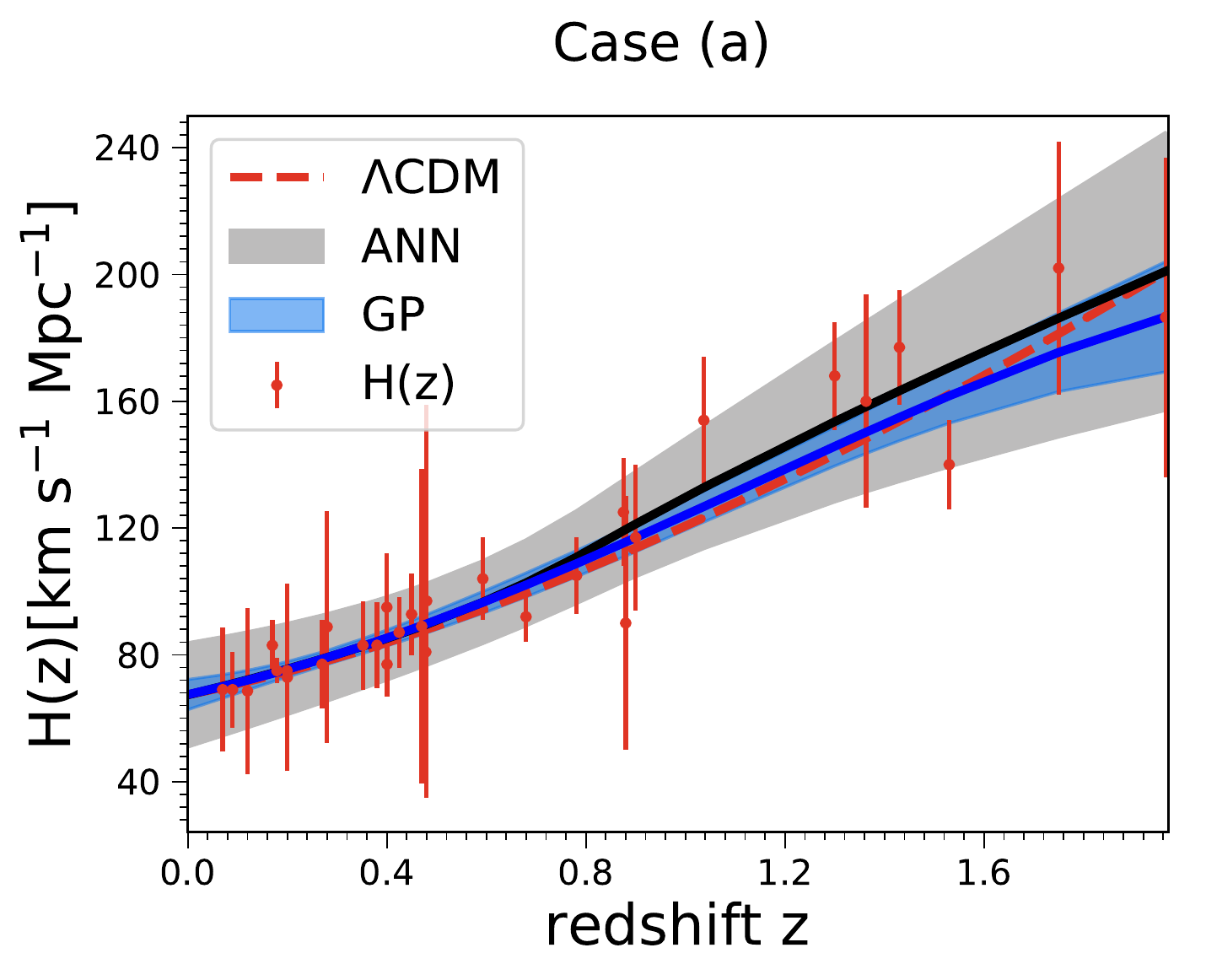}
	\includegraphics[width=0.32\textwidth]{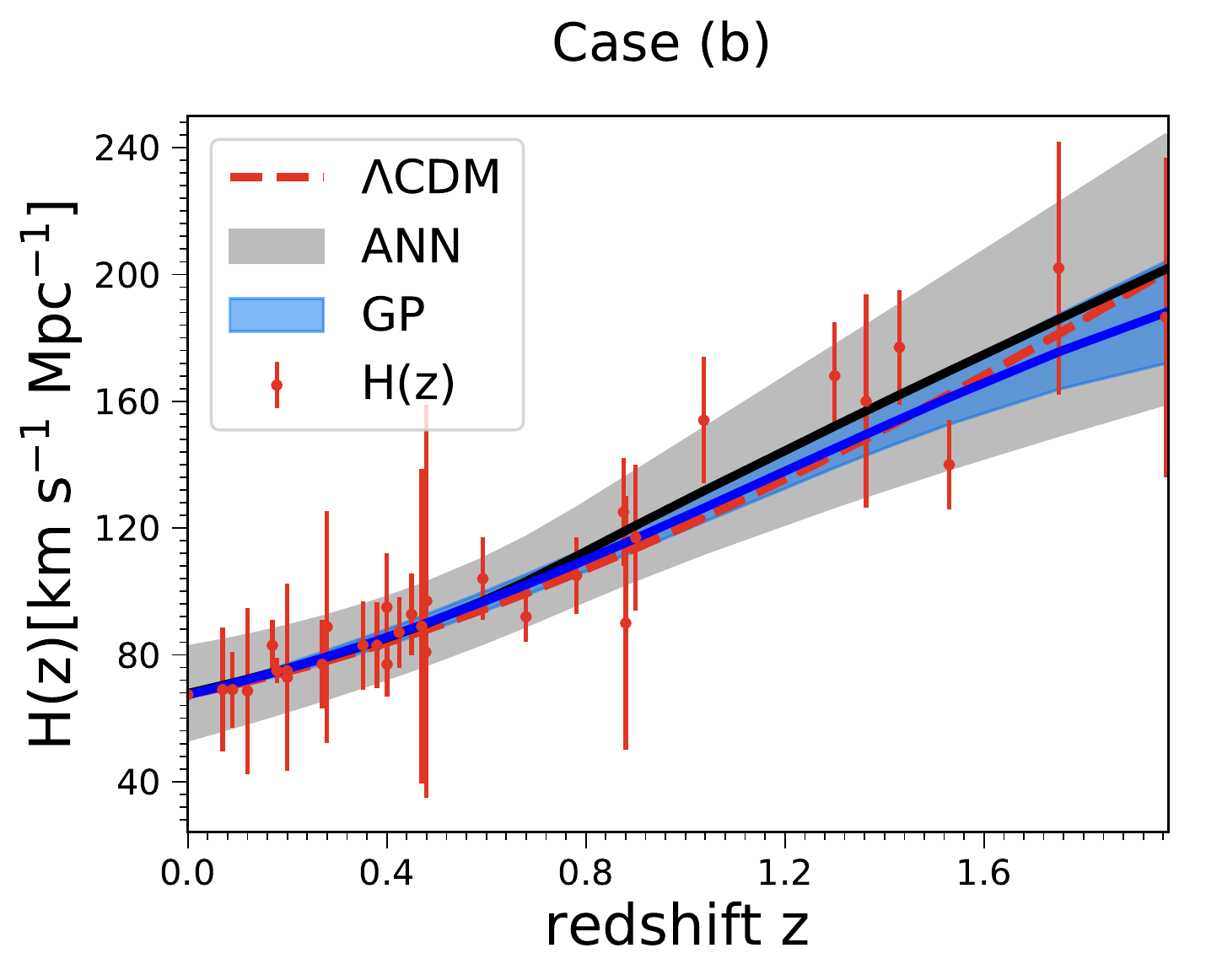}
	\includegraphics[width=0.32\textwidth]{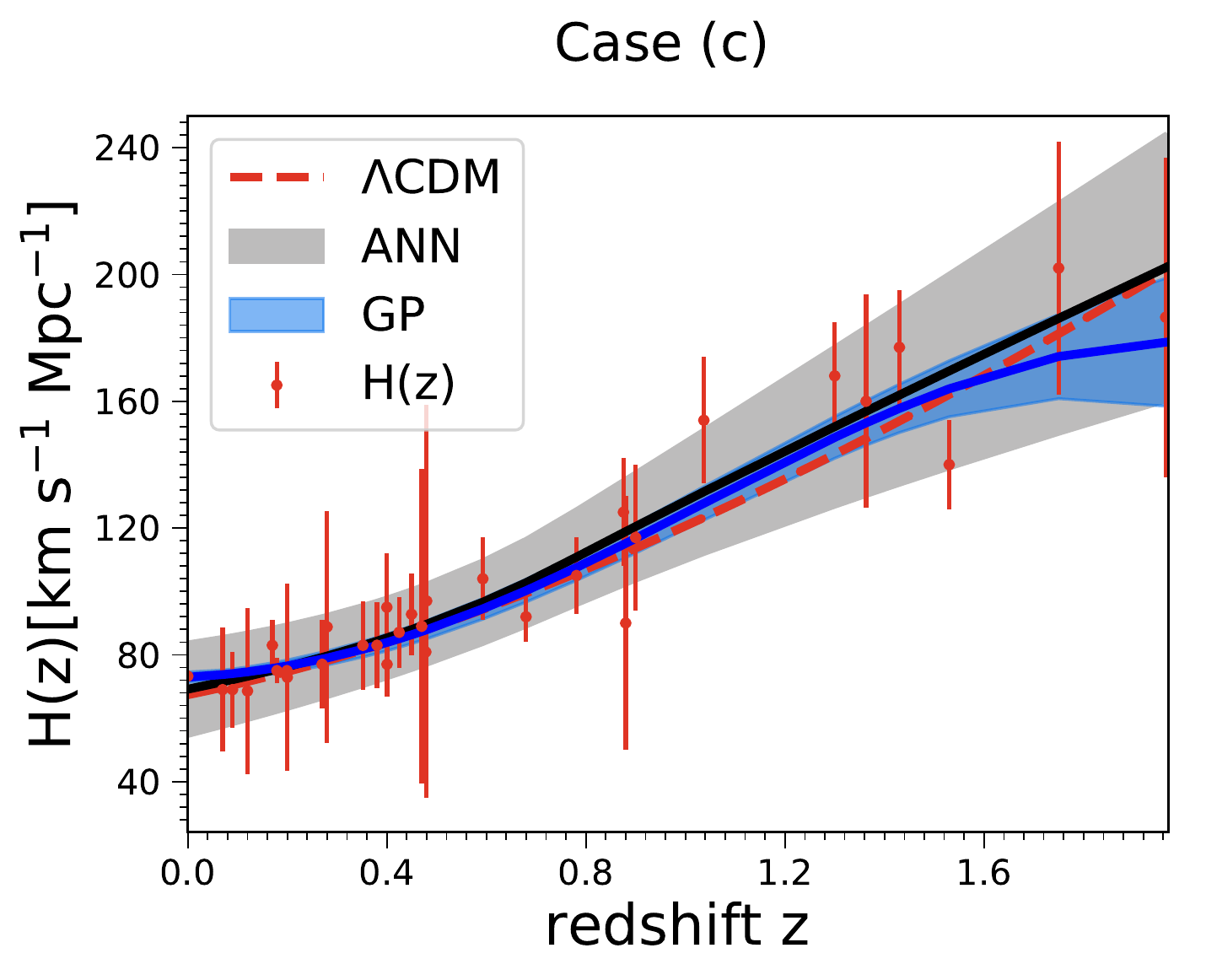}
	\caption{Reconstructed functions of $H(z)$ with ANN and GP. The red dots with error bars represent the $H(z)$ data, while the red dashed lines correspond to the flat $\Lambda$CDM models with $H_0=67.4 ~\rm km\ s^{-1}\ Mpc^{-1}$ and $\Omega_{\rm m}=0.315$ (Planck2018 result).}\label{fig:rec_Hz_ann_gp}
\end{figure*}

\subsection{Hubble parameter $H(z)$}

The Hubble parameter measurements $H(z)$ describe the expansion rate of the universe, and it has been used to explore the evolution of the universe and the nature of dark energy. $H(z)$ can be obtained in two ways. One method is based on the detection of the radial BAO features \citep{Gaztanaga:2009,Blake:2012,Samushia:2013}. However, an fiducial cosmological model is assumed when using this method, thus the $H(z)$ data obtained via this method is model-dependent, which is not suitable for our model-independent analysis. Another method that obtains $H(z)$ is to calculate the differential ages of passively evolving galaxies at different redshifts \citep{Jimenez:2002}, which makes the $H(z)$ measurements are cosmological model-independent and thus can be used in our analysis. In this method, a change rate $\Delta z/\Delta t$ can be obtained, then the Hubble parameter $H(z)$ can be written as
\begin{equation}
H(z)\simeq-\frac{1}{1+z}\frac{\Delta z}{\Delta t} ~.
\end{equation}
This method is usually called the cosmic chronometers and the $H(z)$ data based on this method refers to as CC $H(z)$. The sample of CC $H(z)$ collected in \citet{Wanggj:2020} is taken in our analysis, which has 31 data points in the redshift range of [0.07, 1.965].

\subsection{Function of $H(z)$}\label{sec:reconstruct_Hz}

In our analysis, the cosmic curvature is constrained by comparing the distance modulus from $H(z)$ and that from SNe Ia. It should be noted that for a specific SN Ia data point, there is no corresponding $H(z)$ measurement. One possible way is to reconstruct functions of $H(z)$ to make it possible that there is a $H(z)$ measurement for each SNe Ia at the specific redshift. Therefore, using the ANN and the GP method, we reconstruct a function of the CC $H(z)$ to achieve the model-independent constraint on the cosmic curvature.

It should be noted that the minimum redshift of the CC $H(z)$ is 0.07, which is larger than many of the SNe Ia data. Therefore, if we want to use the Hubble parameter to explore a lower redshift universe, one possible method is to extend the reconstructed function of $H(z)$ to a lower redshift. However, this extension is completely approximate, and when there is a small amount of $H(z)$ data near the redshift interval, bias may be introduced. Therefore, a prior of $H_0$ is considered in the reconstruction of $H(z)$ to make the reconstructed function more reliable. Specifically, we adopt two recent measurements of $H_0$: $H_0=73.24\pm1.74 ~\rm km\ s^{-1}\ Mpc^{-1}$ with 2.4\% uncertainty \citep{Riess:2016}, and  $H_0=67.4\pm0.5 ~\rm km\ s^{-1}\ Mpc^{-1}$ with 0.7\% uncertainty \citep
{Planck2018:VI}. In addition, for comparison purposes, we also reconstruct $H(z)$ with no prior of $H_0$. Therefore, there are three cases for the reconstruction of $H(z)$:
\begin{itemize}
	\item [(a)] with no prior of $H_0$;
	\item [(b)] with a prior of $H_0=67.4\pm0.5 ~\rm km\ s^{-1}\ Mpc^{-1}$; and
	\item [(c)] with a prior of $H_0=73.24\pm1.74 ~\rm km\ s^{-1}\ Mpc^{-1}$.
\end{itemize}
The sample of case (a) has 31 observational $H(z)$ data, and for cases (b) or (c), the sample contains 32 data points.

\subsubsection{Reconstruction with ANN}

When reconstructing a function from data with ReFANN, one should firstly tune hyperparameters of the neural network to find an optimal ANN model \citep{Wanggj:2020}. Here we adopt the optimal ANN model of reconstructing functions from Hubble parameter that is selected by \citet{Wanggj:2020}. The optimal model has one hidden layer, and there are 4096 neurons totally in the hidden layer. Using this model, we reconstruct functions of $H(z)$ for the three $H(z)$ samples, and the results are shown in Figure \ref{fig:rec_Hz_ann_gp} (the black solid lines with gray areas). The red dots with error bars is the CC $H(z)$ and the red dashed line represent the flat $\Lambda$CDM model with $H_0=67.4 ~\rm km\ s^{-1}\ Mpc^{-1}$ and $\Omega_{\rm m}=0.315$ (Planck2018 result, \citet{Planck2018:VI}).

Obviously, the three functions of $H(z)$ are consistent with the flat $\Lambda$CDM model within a $1\sigma$ confidence level. Furthermore, the three functions of $H(z)$ are obviously consistent with each other within a $1\sigma$ confidence level. Specifically, for the best values of the reconstructed functions (the black solid lines in Figure \ref{fig:rec_Hz_ann_gp}), the relative deviation of case (b) with respect to case (a) is $<0.9\%$, and the relative deviation of case (c) with respect to case (a) is $<2.0\%$. These results indicate that the function of $H(z)$ reconstructed by ANN is not sensitive to the prior of $H_0$. Moreover, it should be noted that, for case (a), the reconstructed Hubble constant is
\begin{equation}\label{equ:rec_H0_ann}
H_0=67.35\pm16.47~ \rm km~ s^{-1} ~Mpc^{-1}~,
\end{equation}
which is, for the best-fit value, very similar to the latest {\it Planck} CMB result: $H_0=67.4\pm 0.5~\rm km~ s^{-1} ~Mpc^{-1}$, and is also consistent with that of \citet{Wanggj:2020}.

\subsubsection{Reconstruction with GP}\label{subsec:rec_with_GP}

In addition to the ANN method, we also reconstruct $H(z)$ with GP, and the python package GaPP \citep{Seikel:2012a} is used to execute the Gaussian process in our analysis. GaPP will reconstruct a function for the given data and return the mean and standard deviation of the reconstructed function at a specific redshift \citep{Seikel:2012a}. In the Gaussian process, it is assumed to have correlation between the function value at $x$ and the function value at some other point $\tilde{x}$, which is related by a covariance function. Therefore, when reconstructing functions with GP, it is necessary to select a specific covariance function. Here, we adopt the commonly used squared exponential covariance function:
\begin{equation}\label{equ:GP_kernel}
k(x,\tilde{x}) =
\sigma_f^2 \exp\left( -\frac{(x - \tilde{x})^2}{2\ell^2} \right) \;,
\end{equation}
where $\sigma_f$ and $\ell$ are two hyperparameters that can be optimized. This function is infinitely differentiable and is the default setting of GaPP. We use the {\it dgp} function in GaPP to reconstruct $H(z)$. The two hyperparameters $\sigma_f$ and $\ell$ are automatically optimized in this function.

The reconstructed functions of $H(z)$ for the three cases are shown in Figure \ref{fig:rec_Hz_ann_gp} with blue solid lines and areas. We can see that these $H(z)$ functions are consistent with the flat $\Lambda$CDM model (the red dashed lines) within a $1\sigma$ confidence level for cases (a) and (b), while there is a little deviation for case (c). This indicates that the prior of $H_0$ will greatly influence the reconstruction of $H(z)$ when using GP, which is slightly different from that of the ANN method. The relative deviation of the best values of the reconstructed $H(z)$ (the blue solid lines) is $< 2.9\%$ for cases (a) and (b), and $< 15.3\%$ for cases (a) and (c) (this is quite large). Moreover, the reconstructed Hubble constant for case (a) is
\begin{equation}\label{equ:rec_H0_gp}
H_0=67.4\pm4.75~ \rm km~ s^{-1} ~Mpc^{-1}~,
\end{equation}
where the best-fit value is very similar to the result of the ANN method (Equation \ref{equ:rec_H0_ann}), and also similar to the latest {\it Planck} CMB result.

\section{Cosmic curvature}\label{sec:constraint_omk}

In this section, we achieve the constraint on the cosmic curvature $\Omega_K$ by comparing the distance modulus of $H(z)$ and that obtained from the SNe Ia or GW. In section \ref{sec:omk_from_Hz_SNe}, we introduce the constraint on $\Omega_K$ using $H(z)$ and SNe Ia, and in section \ref{sec:omk_from_Hz_GW} we show the result constrained from $H(z)$ and GW.

\subsection{Cosmic curvature from $H(z)$ \& SNe Ia}\label{sec:omk_from_Hz_SNe}

\subsubsection{SNe Ia data}

The sample of SNe Ia used in our analysis is the latest Pantheon SNe Ia \citep{Scolnic:2018}, which contains 1048 data points within the redshift range of [0.01, 2.26]. For these SNe Ia, two nuisance parameters, $\alpha$ and $\beta$, are recovered by using the BEAMS with Bias Corrections (BBC) method \citep{Kessler:2017}, and the corrected apparent magnitudes $m_{B,corr}^* = m_{B}^*+\alpha\times x_1-\beta\times c + \Delta_B$ for all the SNe Ia are reported in \citep{Scolnic:2018}, where $\Delta_B$ is a distance correction based on predicted biases from simulations. Therefore, the distance modulus of Pantheon SNe Ia can be rewritten as
\begin{equation}
\mu=m_{B,corr}^* - M_{B} ~,
\end{equation}
where $M_B$ represents the absolute magnitude of the $B$ band and it should be constrained simultaneously with the cosmological parameters. In the Pantheon sample, there is only one SNe (the SNe whose redshift is 2.26) exceeds the redshift range of the reconstructed function of $H(z)$. Thus, this SNe is not considered in our analysis, which means that there are only 1047 SNe Ia are taken in our analysis.

\subsubsection{Method}\label{sec:omk_from_Hz_SNe_method}

In section \ref{sec:reconstruct_Hz}, we reconstruct functions of CC $H(z)$ using the ANN and GP methods. Then, the total line-of-sight comoving distance $D_{C}$ \citep{Hogg:1999} can be further derived from the reconstructed functions of $H(z)$ by using
\begin{equation}\label{equ:comoving}
D_{C}=c\int_{0}^{z}\frac{dz'}{H(z')} ~,
\end{equation}
where the error of $D_C$ is obtained by integrating the error of the function of $H(z)$. Furthermore, the luminosity distance $D_L$ can be obtained from $D_{C}$ via
\begin{equation}\label{equ:dl_Hz}
\frac{D_{L}}{(1+z)}=
\begin{cases}
\frac{D_{H}}{\sqrt{\Omega_K}} \sinh{[\sqrt{\Omega_K}D_{C}/D_{H}]}&\Omega_K>0\\D_{C} &\Omega_K=0\\ \frac{D_{H}}{\sqrt{\left| \Omega_K \right|}} \sin{[\sqrt{\left| \Omega_K \right|}D_{C}/D_{H}]}&\Omega_K<0 ~,
\end{cases}
\end{equation}
where $D_{\rm H}=cH_0^{-1}$. The reconstructed distance modulus $\mu_H$ can be further obtained by using 
\begin{align}\label{equ:mu_LCDM}
\mu_H&=5\log\frac{D_L}{\rm Mpc}+25 ~, & D_L&=(1+z)D_C ~,
\end{align}
and the corresponding errors can be propagated by using
\begin{align}
\sigma_{\mu_H}&=\frac{5}{\ln10}\frac{\sigma_{D_L}}{D_L}~,  & \sigma_{D_L}&= (1+z)\sigma_{D_C} ~.
\end{align}
Finally, the cosmic curvature $\Omega_K$ can be constrained by minimizing the $\chi^{2}$ statistic,
\begin{equation}\label{equ:chi2_SNe}
\chi^{2}=\bm{\Delta}\hat{\mu}^{T}\cdot \bm{Cov}^{-1}\cdot \bm{\Delta}\hat{\mu} ~,
\end{equation}
where $\bm{Cov}$ is the full covariance matrix, and $\Delta\hat{\mu}=\hat{\mu}_{\rm SNe}-\hat{\mu}_H$ is the difference between the distance modulus of SNe Ia and that of the $H(z)$ data. For the Pantheon SNe Ia, the BBC method produces distances from the fit parameters directly, thus, there is only a single systematic covariance matrix $\bm{C}_{sys}$.

\subsubsection{$\Omega_K$ from ANN}\label{sec:omk_from_Hz_SNe_ann}

\begin{figure*}
	\centering
	\includegraphics[width=0.45\textwidth]{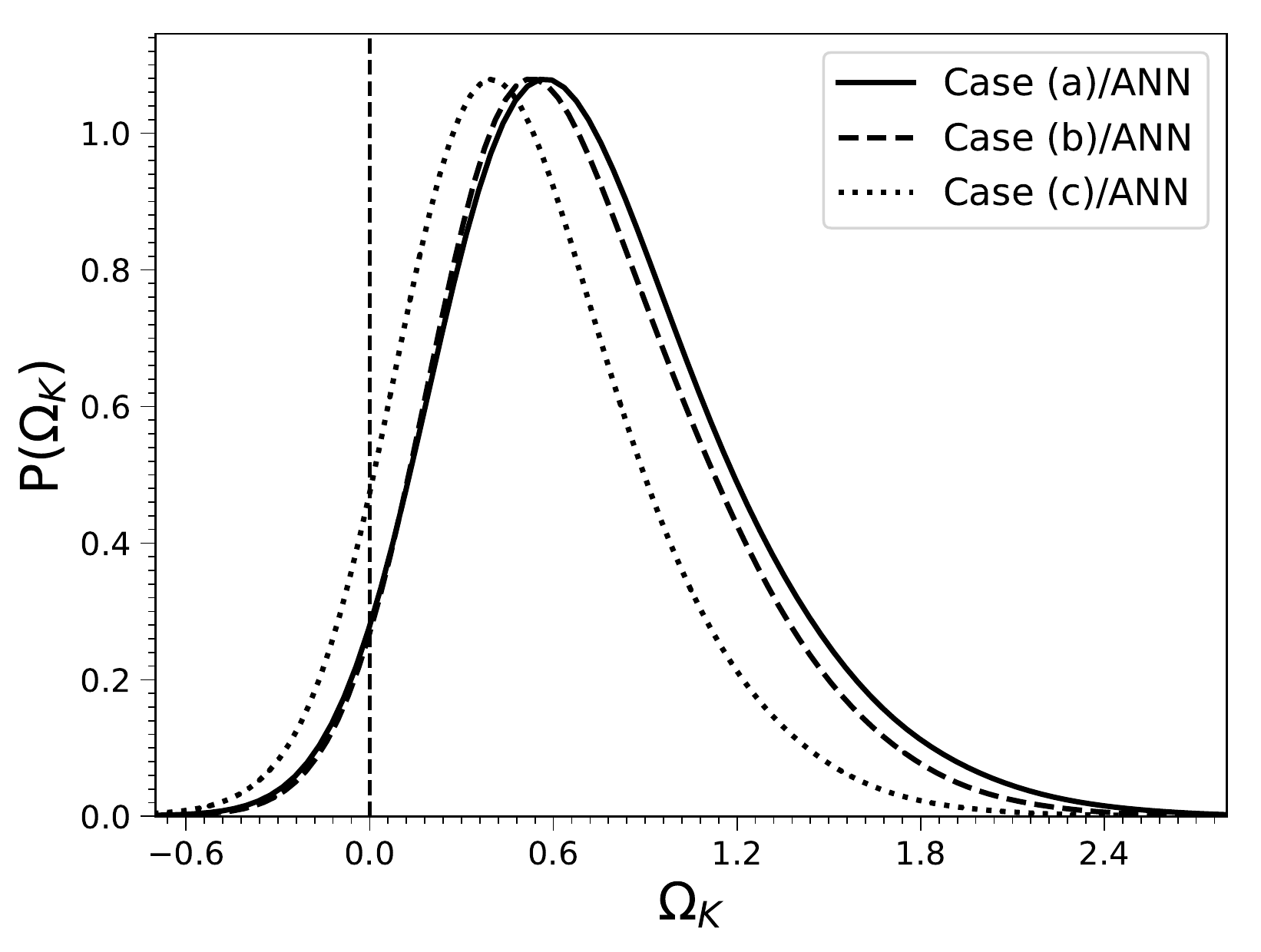}
	\includegraphics[width=0.45\textwidth]{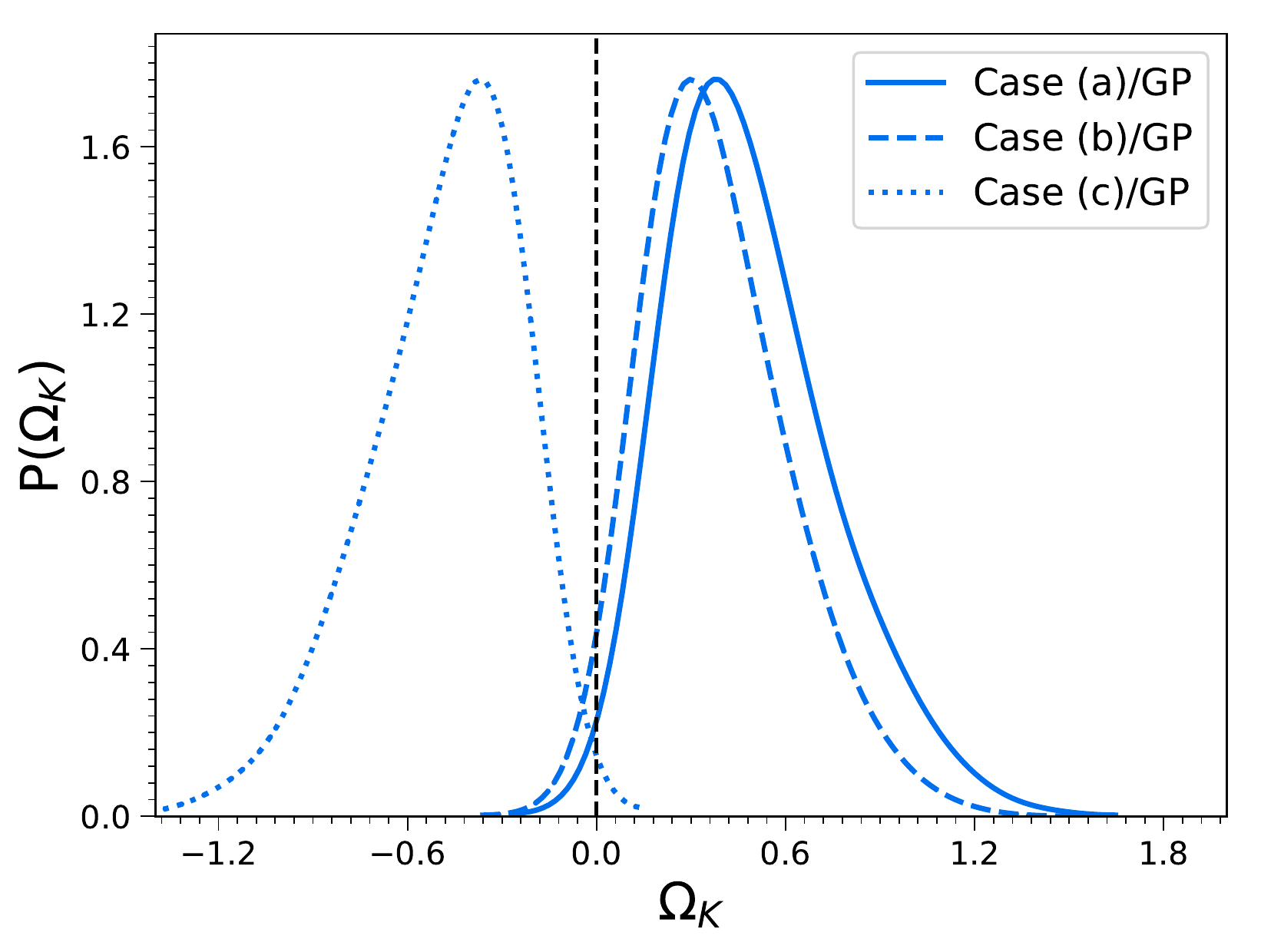}
	\caption{One-dimensional marginalized distribution of $\Omega_K$ constrained from $H(z)$ and Pantheon SNe Ia. {\it Left:} the $H(z)$ is reconstructed by using the ANN method. {\it Right:} the $H(z)$ is reconstructed by using the GP method.}\label{fig:omk_ann_gp}
\end{figure*}

\begin{table}
	\centering\caption{$1\sigma$ Constraints on $\Omega_K$ constrained from $H(z)$ and Pantheon SNe Ia. The $H(z)$ is reconstructed by using the ANN method. Case (a), (b) and (c) represent no prior of $H_0$, $H_0 = 67.74\pm 0.46~\rm km~ s^{-1} ~Mpc^{-1}$, and $H_0 = 73.24\pm 1.74~\rm km~ s^{-1} ~Mpc^{-1}$ when reconstructing the function of $H(z)$, respectively. See the text for details.}\label{tab:omk_ann}
	\begin{tabular}{c|c|c|c}
		\hline\hline
		Cases & (a) & (b) & (c) \\
		\hline
		$\Omega_K$ & $0.665\pm0.419$ & $0.626\pm0.340$ & $0.461\pm0.349$ \\
		\hline\hline
	\end{tabular}
\end{table}

Using the Markov chain Monte Carlo (MCMC) method, the best-fit value and uncertainty of the cosmic curvature can be obtained by generating sample points of the probability distribution to $\Omega_K$ and other nuisance parameters simultaneously. One-dimensional distribution of $\Omega_K$ constrained from $H(z)$+Pantheon is shown in the left panel of Figure \ref{fig:omk_ann_gp}, and the corresponding best-fit value with $1\sigma$ errors are listed in Table \ref{tab:omk_ann}. The black solid, dashed and dotted lines referring to the result of case (a), (b) and (c), respectively. We can see that the results of cases (b) and (c) are a little different from that of the case (a), which should be caused by the prior of $H_0$ in the reconstruction of $H(z)$. However, the results of these three cases are consistent with each other within a $1\sigma$ confidence level, and all of them favor a positive value of $\Omega_K$, and the differences between these three results and the value inferred from Planck CMB are $1.6\sigma$, $1.8\sigma$, and $1.3\sigma$, respectively. Therefore, a spatially open universe may be preferred by $H(z)$ and Pantheon SNe Ia.

In addition, we can see that the best-fit value of case (a) is similar to that of case (b), while for case (c), the best-fit value is a little smaller than that of cases (a) and (b). This is not difficult to understand, because the reconstructed Hubble constant (Equation \ref{equ:rec_H0_ann}) of the case (a) is similar to the prior of $H_0$ in case (b). Furthermore, the consistency of the results of these three cases within a $1\sigma$ confidence level may indicate that the ANN method is capable to extend the reconstructed $H(z)$ function to a lower redshift to explore a lower redshift universe. Thus, the prior of the Hubble constant does not have a significant effect on the estimation of the cosmic curvature.

\subsubsection{$\Omega_K$ from GP}\label{sec:omk_from_Hz_SNe_gp}

Following the same procedure of section \ref{sec:omk_from_Hz_SNe_method}, the total line-of-sight comoving distance $D_C$ can be derived from the functions of $H(z)$ reconstructed by GP, then the the luminosity distance $D_L$ can be further obtained from $D_C$, and finally the distance modulus can be calculated by using $D_L$. Then, the cosmic curvature $\Omega_K$ can be constrained by minimizing the $\chi^2$ statistic of Equation \ref{equ:chi2_SNe}. 

\begin{table}
	\centering\caption{The same as Table \ref{tab:omk_ann}, but now the $H(z)$ is reconstructed by using the GP method.}\label{tab:omk_gp}
	\begin{tabular}{c|c|c|c}
		\hline\hline
		Cases & (a) & (b) & (c) \\
		\hline
		$\Omega_K$ & $0.447\pm0.248$ & $0.348\pm0.217$ & $-0.440\pm0.234$ \\
		\hline\hline
	\end{tabular}
\end{table}

The cosmic curvature constrained from $H(z)$+Pantheon are listed in Table \ref{tab:omk_gp} and the one-dimensional marginalized distribution of $\Omega_K$ is shown in the right panel of Figure \ref{fig:omk_ann_gp}. We can see the result of the case (a) is similar to that of the case (b), and both them favor a positive value of $\Omega_K$, and the differences of the results and that inferred from the Planck CMB are $1.8\sigma$ and $1.6\sigma$. However, the result of the case (c) favors a negative value of $\Omega_K$ and the difference of the result and that of Planck CMB is $1.9\sigma$, which is statistically quite different from that of cases (a) and (b). These results indicate that the prior of $H_0$ will greatly influence the measurement of the cosmic curvature when using GP. 

Moreover, we can see that the results of cases (a) and (b) are consistent with that obtained using the ANN method within a $1\sigma$ confidence level. However, it is quite different for case (c), and the difference between the result and that based on the ANN method is $2.1\sigma$. Therefore, comparing the results of sections \ref{sec:omk_from_Hz_SNe_ann}, we can see that the cosmic curvature based on the ANN method is more stable than that based on the GP method, which may indicate that the ANN method will surpass the GP method in the measurement of the cosmic curvature.

\subsection{Cosmic curvature from $H(z)$ \& GW}\label{sec:omk_from_Hz_GW}

The positive value of $\Omega_K$ preferred by the current CC $H(z)$ and SNe Ia data is contrary to the flat universe supported by CMB experiments in $\Lambda$CDM model. Any deviation from $\Omega_K=0$ would have a profound impact on inflation models and fundamental physics. Thus, this result should be taken seriously and tested strictly. Therefore, in order to further test the reliability of the ANN method, and the potentiality of GW in the estimation of the cosmic curvature, we constrain $\Omega_K$ using the simulated Hubble parameter and GW standard sirens in a model-independent way. We first introduce the simulation of GW standard sirens and then achieve the estimation of the cosmic curvature.

\subsubsection{GW standard sirens}

The chirping GW signals from inspiralling compact binaries can provide an absolute measure of the luminosity distance because the amplitude of GW depends on the so-called chirp mass and the luminosity distance. The chirp mass is measured from the phasing of GW, thus, the luminosity distance $D_L$ can be extracted from the amplitude, which makes GW known as standard siren \citep{Schutz:1986,Abbott:2017}. Besides, if the compact binaries are black hole-neutron star (BH-NS) or binary neutron stars (NS-NS), the redshift of GW sources can be obtained from its electromagnetic counterpart (EM). Therefore, this offers a model-independent way to establish the $D_L-z$ relation over a wide range of redshift and thus can be used in our model-independent estimation of the cosmic curvature.

The GW events used here is simulated from the Einstein Telescope (ET, \citet{ET:2011}), a third generation gravitational wave detector that is designed with high-sensitivity and wide frequency range ($1-10^4$Hz), and would be able to detect the NS-NS mergers up to the redshift of $z\sim2$ and BH-NS mergers up to $z>2$ \citep{Punturo:2010}. We firstly generate the redshift of the GW events. The redshift distribution of GW events is taken to has the form of \citep{Zhao:2011}
\begin{equation}
P(z)\propto \frac{4\upi D_C^2(z)R(z)}{H(z)(1+z)},
\end{equation}
where $D_C$ is the total line-of-sight comoving distance and $R(z)$ describes the time evolution of the burst rate with the form \citep{Schneider:2001,Cutler:2009}
\begin{equation}
R(z)=\begin{cases}
1+2z, & z\leq 1 \\
\frac{3}{4}(5-z), & 1<z<5 \\
0, & z\geq 5.
\end{cases}
\end{equation}
The redshift of GW events is simulated according to this distribution. Then we calculate the luminosity distance in a flat $\Lambda$CDM model using
\begin{equation}\label{equ:DL_LCDM}
D_L(z)=\frac{c(1+z)}{H_0}\int_0^z\frac{dz}{\sqrt{\Omega_{\rm m}(1+z)^3 + \Omega_K(1+z)^2 + \Omega_\Lambda}}~,
\end{equation}
where $\Omega_\Lambda=1-\Omega_{\rm m}-\Omega_K$ and the fiducial $H_0=70~\rm km\ s^{-1}\ Mpc^{-1}$, $\Omega_{\rm m}=0.3$ and $\Omega_K=0$.

\begin{figure}
	\centering
	\includegraphics[width=0.45\textwidth]{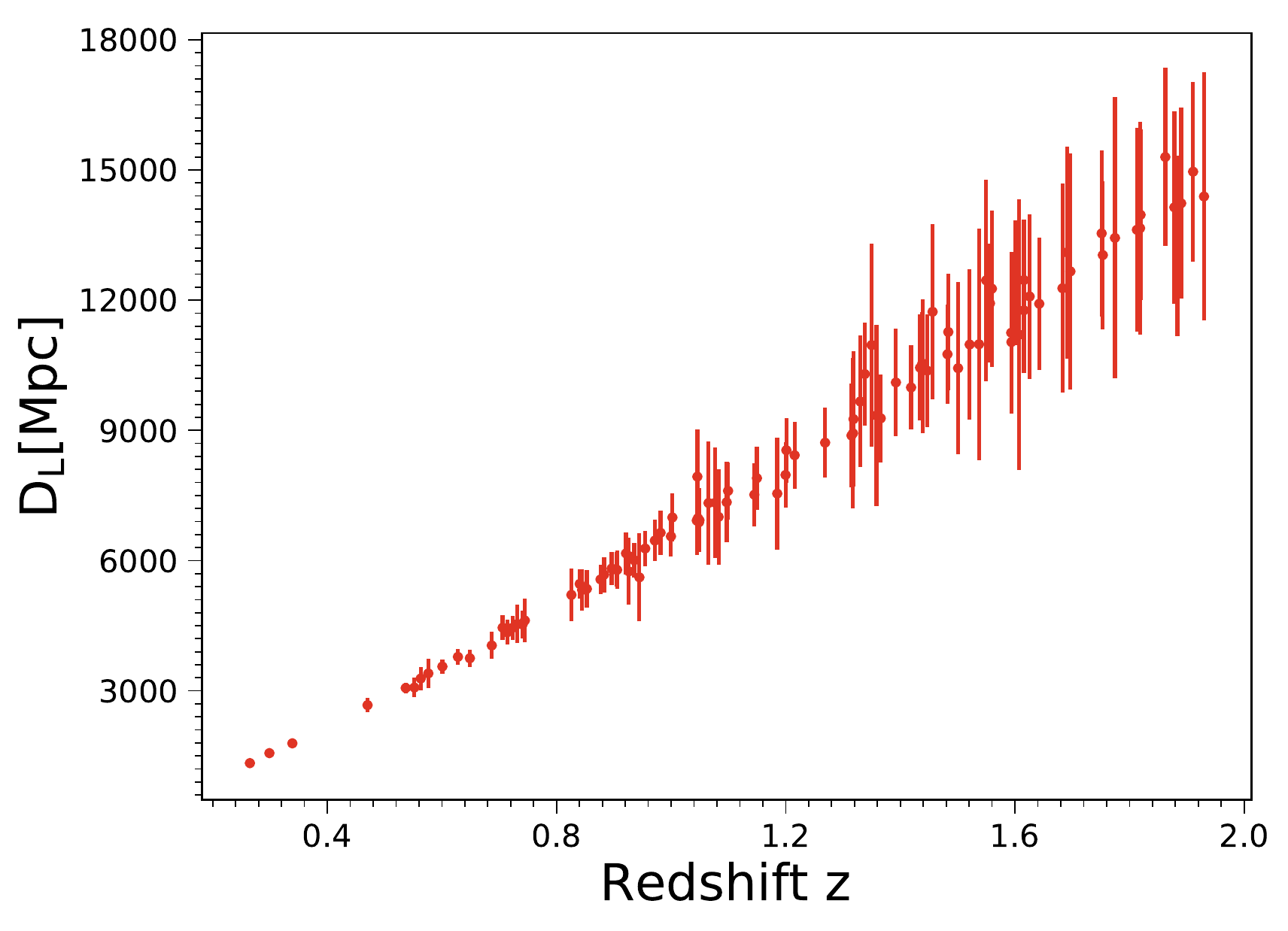}
	\caption{The simulated luminosity distances of 100 GW events observed by ET.}\label{fig:sim_gw100_DL}
\end{figure}

Then the uncertainty of the luminosity distance $\sigma_{D_L}$ of GW events is simulated by following the simulation process of \cite{Cai:2017}. The total uncertainty of luminosity is
\begin{equation}
\sigma_{D_L}=\sqrt{(\sigma_{D_L}^{\rm inst})^2+(\sigma_{D_L}^{\rm lens})^2}
\end{equation}
where $\sigma_{D_L}^{\rm inst}$ is the instrumental error of the luminosity distance, and $\sigma_{D_L}^{\rm lens}$ is the additional error due to the weak lensing which is assumed to be $\sigma_{D_L}^{\rm lens}/D_L=0.05z$. For the simulation of $\sigma_{D_L}^{\rm inst}$, we refer the reader to \citet{Cai:2017} for the detailed process (see also \citet{Zhao:2011,Wei:2018,Qi:2019b}). The mass distribution is chosen uniformly in the interval $[1,2]M_{\sun}$ for neutron stars and $[3,10]M_{\sun}$ for black holes. As the argument in \citep{Cai:2017}, the ET is expected to detect $\mathcal{O}(10^2)$ GW events with EM counterparts per year, and the ratio of possibly detecting BH-NS and NS-NS events is assumed to be 0.03. Thus we firstly simulate 100 GW events, shown in Figure \ref{fig:sim_gw100_DL}. Note that the maximum redshift of the observational $H(z)$ is $\sim2$, therefore, we only simulate the GW events in $0<z<2$.

\subsubsection{Constraint on $\Omega_K$}\label{sec:omk_from_Hz_GW_results}

In our analysis, 100 $H(z)$ is simulated with the same fiducial model (Equation \ref{equ:DL_LCDM}) by using the method illustrated in \citet{Wanggj:2020}. Using the mock data of $H(z)$ and GW events, we can test the capability of GW to the constraint on the cosmic curvature, as well as the reliability of the ANN method. With the simulated luminosity distance of GW events, we can further obtain the distance modulus $\mu_{GW}$ using Equation \ref{equ:mu_LCDM} and the corresponding errors can be propagated from that of $D_L$ by using
\begin{equation}
\sigma_{\mu_{GW}} = \frac{5}{\ln10}\frac{\sigma_{D_L}}{D_L}~.
\end{equation}
Then we can constrain the cosmic curvature $\Omega_K$ by minimize the $\chi^2$
\begin{equation}\label{equ:chi2_gw_Hz}
\chi^2(\Omega_K) = \sum_{i}\frac{\left[\mu_{H}(z_i;\Omega_K)-\mu_{GW}(z_i)\right]^{2}}{\sigma^2_{\mu_H,i}+\sigma^2_{\mu_{GW,i}}}~.
\end{equation}

\begin{figure}
	\centering
	\includegraphics[width=0.45\textwidth]{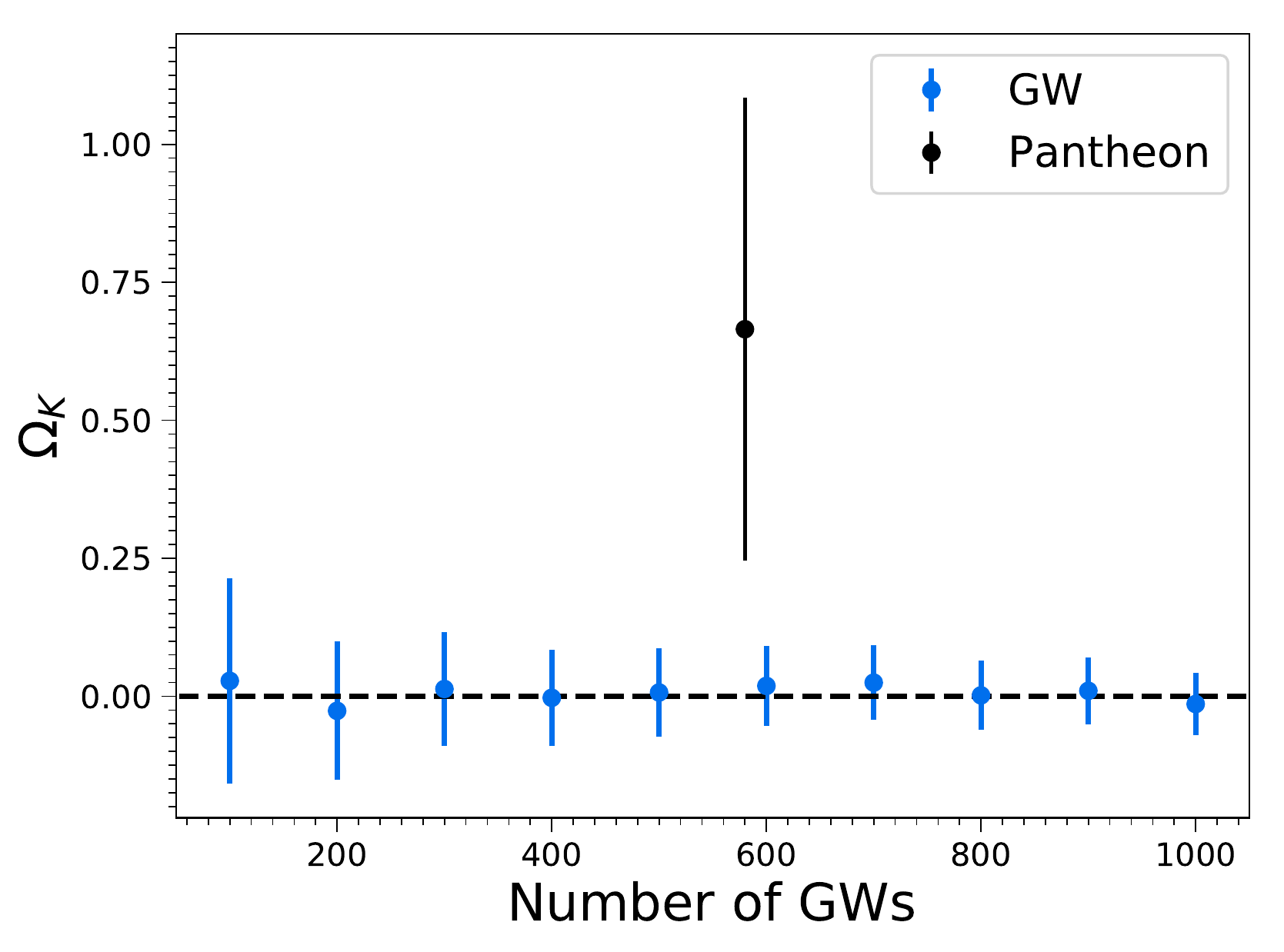}
	\caption{Best-fit cosmic curvature $\Omega_K$ and $1\sigma$ confidence level as a function of the number of GW events. The black dashed line represents the fiducial flat universe.}\label{fig:gw_Hz_omk}
\end{figure}

Specifically, we firstly reconstruct a function of $H(z)$ using the ANN method, then integrate it to obtain the modulus distance $\mu_H$, and finally achieve the estimation of the cosmic curvature using Equation \ref{equ:chi2_gw_Hz}. The cosmic curvature constrained from the simulated $H(z)$ and 100 GW events is
\begin{eqnarray}
\Omega_K = 0.028\pm0.186.
\end{eqnarray}
This result is consistent with the fiducial value of $\Omega_K=0$ within a $1\sigma$ confidence level, which implies the reliability of the ANN method. Moreover, we note that the error of this result is about $56\%$ smaller than that constrained using Pantheon SNe Ia (case (a) in Table \ref{tab:omk_ann}). Therefore, the GW standard siren will be a powerful tool in the constraint on the cosmic curvature.

In order to show how effective of the GW standard siren in the constraint on $\Omega_K$, we further simulate several catalogs of GW events with the number varies from 200 to 1000 and use them to constrain the cosmic curvature. The best-fit values of $\Omega_K$ and the corresponding errors as a function of the number of GW events are shown in Figure \ref{fig:gw_Hz_omk}. We can see that the error of $\Omega_K$ decrease with the increase in the number of GW events and all these results are consistent with the fiducial flat universe within a $1\sigma$ confidence level. In addition, the error of $\Omega_K$ will be $0.056$ when 1000 GW events are detected, which is much smaller than that obtained from SNe Ia. Therefore, the Hubble parameter and future gravitational wave standard siren will constrain the cosmic curvature $\Omega_K$ in a model-independent way with high precision.

For the ten sets of constraint on $\Omega_K$, the average relative deviation from the fiducial flat universe is $0.07\sigma$, which is quite small. For comparison, we also reconstruct the Hubble parameter with the GP method, and obtain another ten sets of constraint on $\Omega_K$. The corresponding average relative deviation from the fiducial flat universe is $0.52\sigma$. This deviation is much larger than that based on ANN. Therefore, this indicates that the ANN method will surpass the GP method in the measurement of the cosmic curvature.

\section{Discussions}\label{sec:discussion}

\begin{figure}
	\centering
	\includegraphics[width=0.45\textwidth]{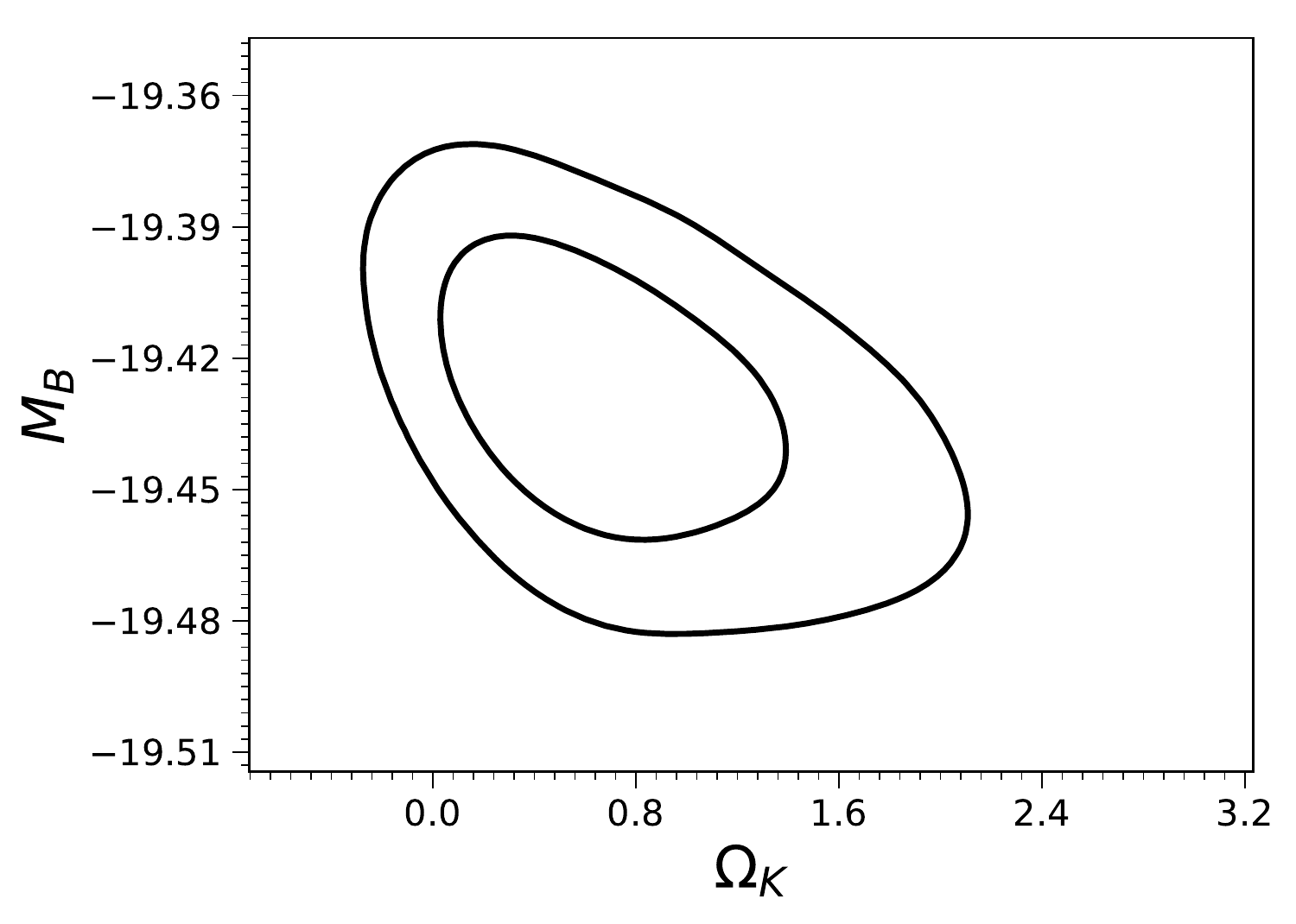}
	\caption{Two-dimensional marginalized distribution for $\Omega_K$ and $M_B$.}\label{fig:omk_mb}
\end{figure}

In section \ref{sec:omk_from_Hz_SNe_ann}, with the function of $H(z)$ reconstructed by the ANN method, we show that a positive value of $\Omega_K$ is favored by the current CC $H(z)$ and Pantheon SNe Ia. The analysis of \citet{Wanggj:2020} and section \ref{sec:omk_from_Hz_GW_results} show that the ANN method is reliable and unbiased for the reconstructed functions of $H(z)$, thus, the constraint on the cosmic curvature should be reliable. However, it should be noted that the cosmic curvature $\Omega_K$ is strongly degenerate with the absolute magnitude $M_B$ of SNe Ia (see Figure \ref{fig:omk_mb}), which means that a small deviation of the absolute magnitude will lead to a great change in the cosmic curvature. Thus, this degeneracy may make it difficult to constrain the cosmic curvature and the nuisance parameters of SNe Ia simultaneously and prevents us from understanding the nature of the cosmic curvature. Moreover, the Hubble parameters used in this work have only 31 observations, which is much smaller than that of SNe Ia. Thus, the number of the current CC $H(z)$ may not be sufficient to represent the actual situation of the Hubble parameters. Therefore, further analysis is needed with more Hubble parameter observations.

\subsection{Effect of covariance functions}

\begin{figure*}
	\centering
	\includegraphics[width=0.45\textwidth]{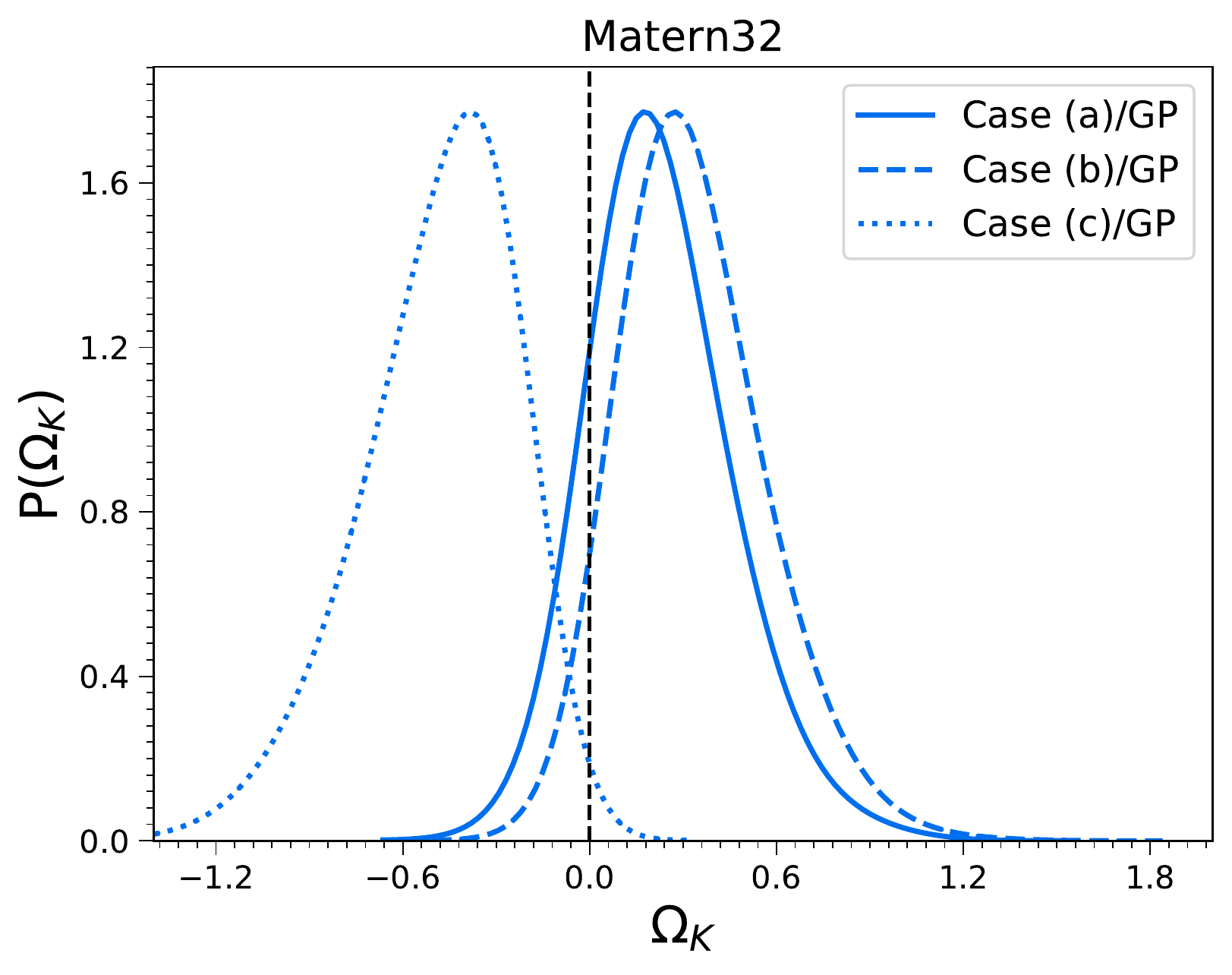}
	\includegraphics[width=0.45\textwidth]{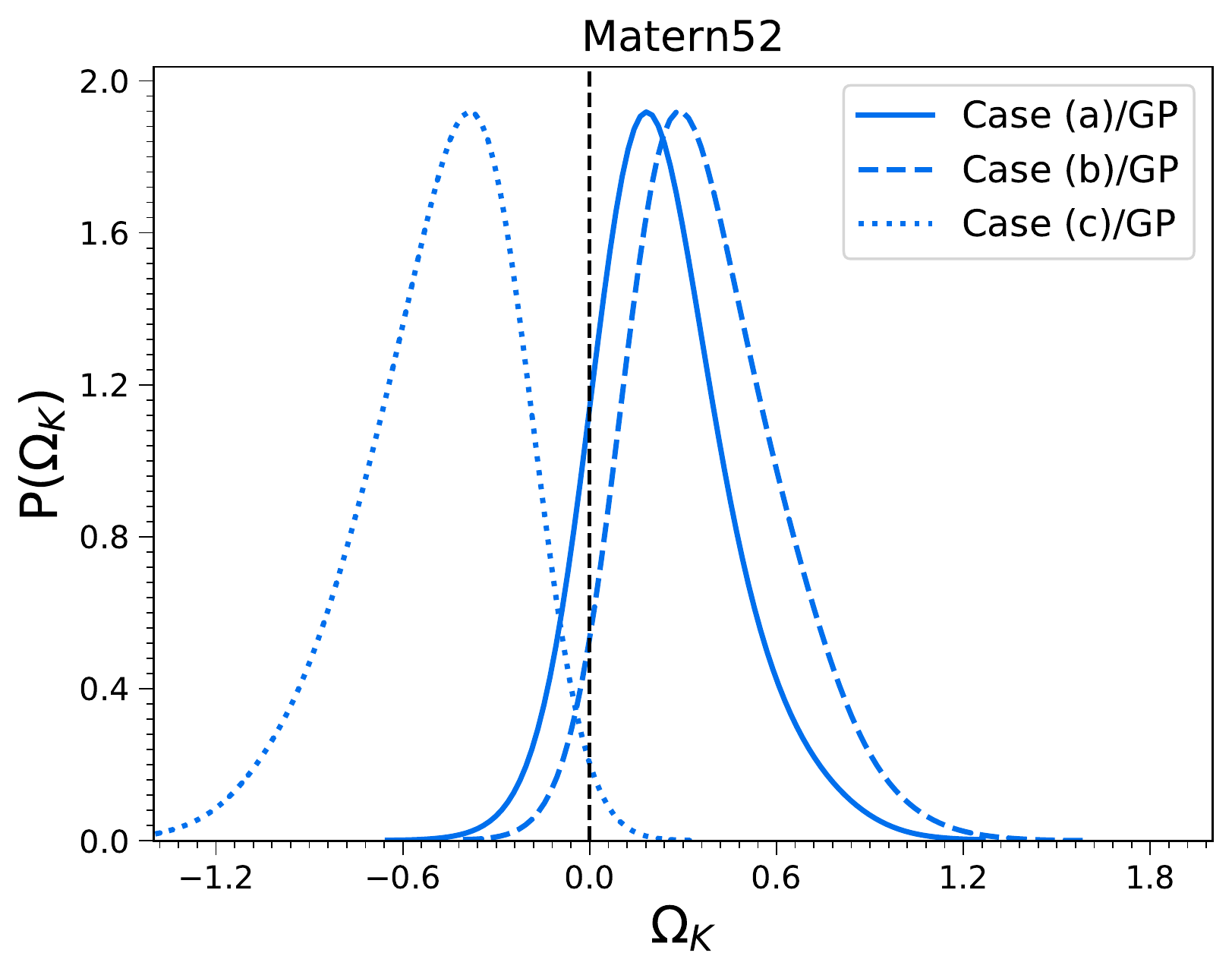}
	\includegraphics[width=0.45\textwidth]{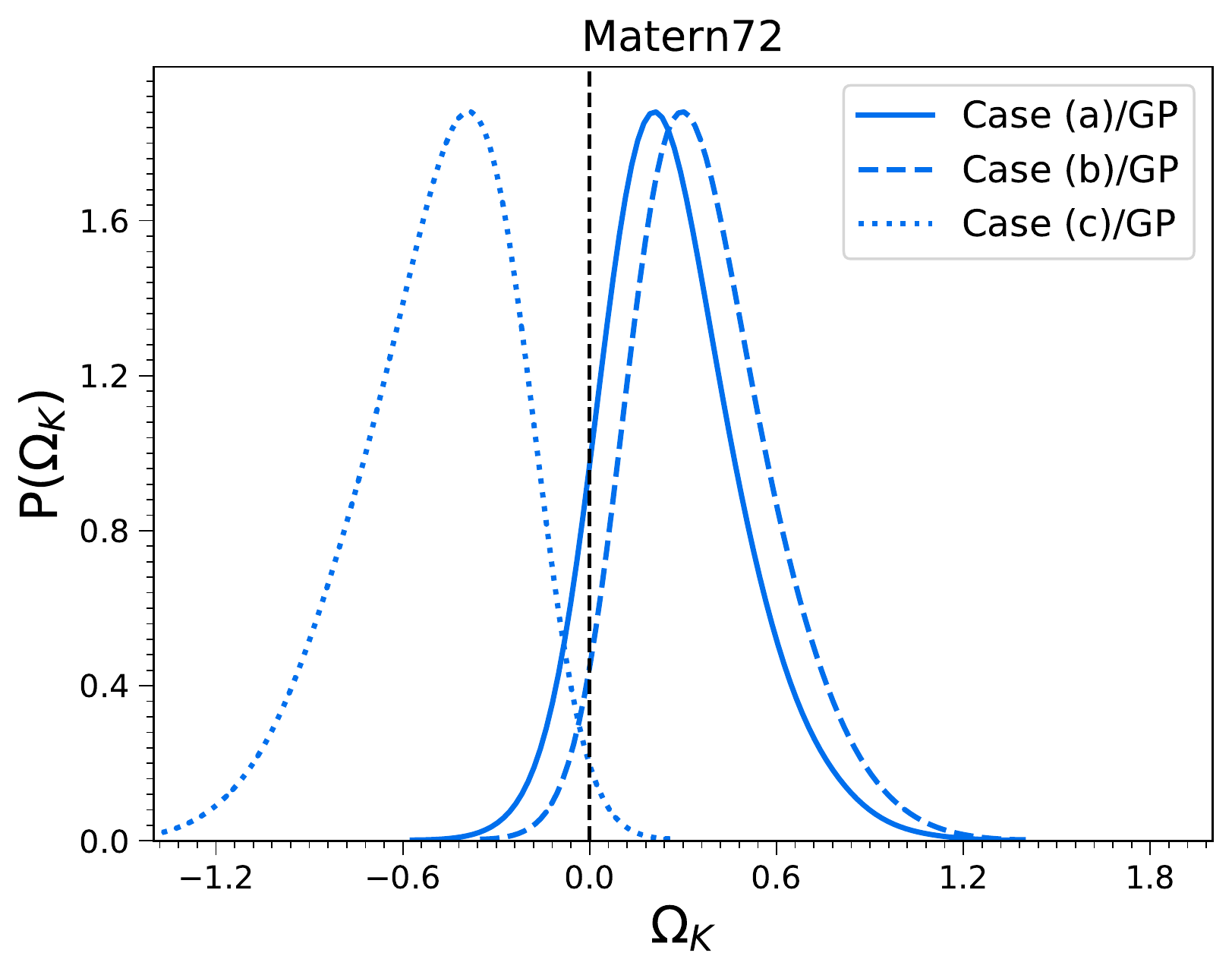}
	\includegraphics[width=0.45\textwidth]{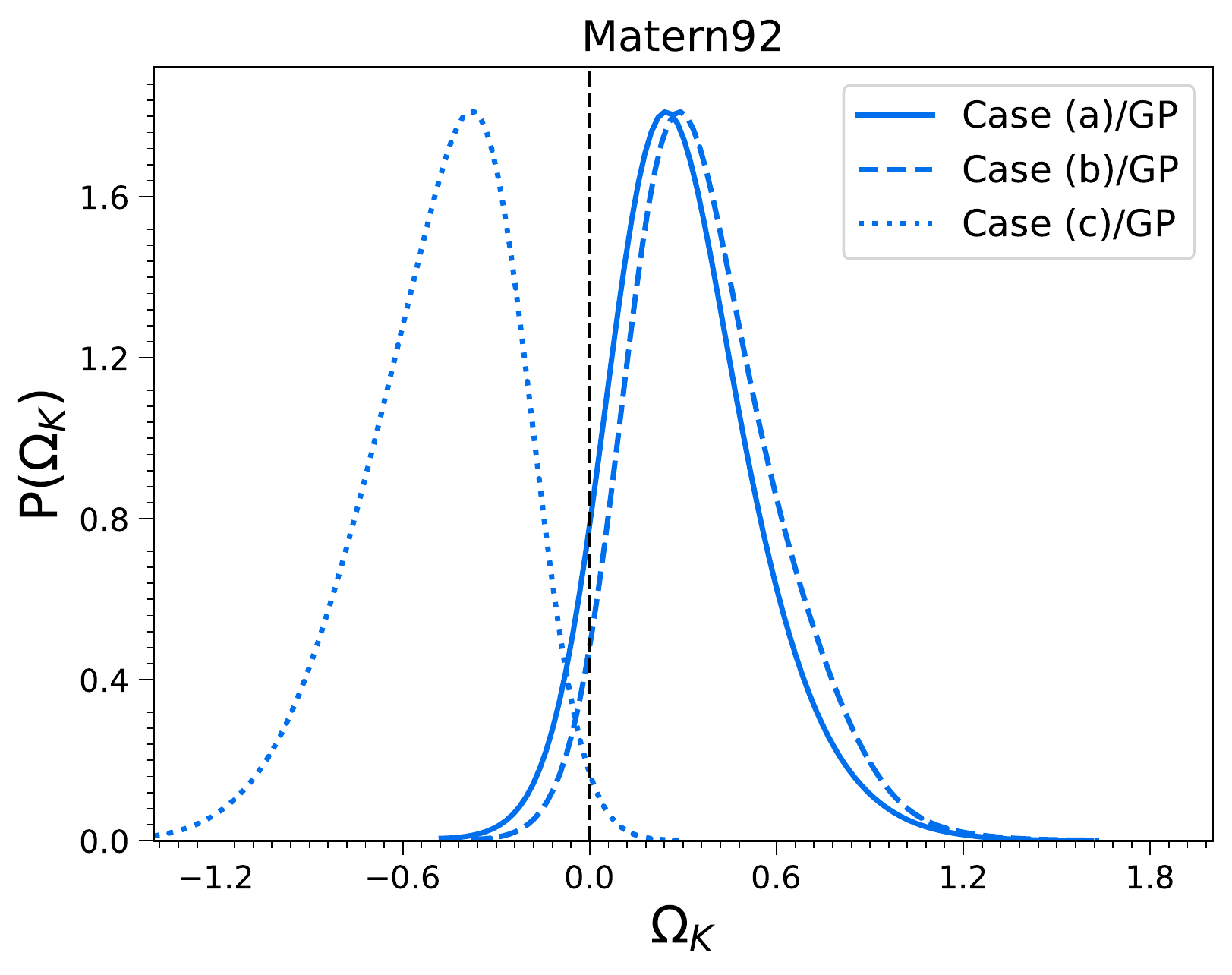}
	\caption{One-dimensional marginalized distribution of $\Omega_K$ constrained from $H(z)$ and Pantheon SNe Ia. The $H(z)$ is reconstructed by using the GP method with different kinds of Matern covariance function.}\label{fig:omk_gp_multiCov}
\end{figure*}

In section \ref{subsec:rec_with_GP}, we reconstructed $H(z)$ using GP, and in section  \ref{sec:omk_from_Hz_SNe_gp} we found the sensitivity of GP to the prior of $H_0$. It should be noted that only the commonly used squared exponential covariance function is adopted in the analysis above. To further illustrate this problem, here we examine the effect of covariance functions on the results. Specifically, we adopt four kinds of Matern class of covariance functions in GaPP: Matern32, Matern52, Matern72, and Matern92, respectively. For Matern32, the covariance function has the form of
\begin{equation}\label{equ:GP_covf_Matern32}
k(x,\tilde{x}) =
\sigma_f^2 \exp\left[-\frac{\sqrt{3}|x - \tilde{x}|}{\ell} \right] \left(1+\frac{\sqrt{3}|x - \tilde{x}|}{\ell} \right),
\end{equation}
for Matern52, the covariance function is
\begin{equation}\label{equ:GP_covf_Matern52}
k(x,\tilde{x}) =
\sigma_f^2 \exp\left[-\frac{\sqrt{5}|x - \tilde{x}|}{\ell} \right] \left(1+\frac{\sqrt{5}|x - \tilde{x}|}{\ell} +\frac{5(x - \tilde{x})^2}{3\ell^2} \right),
\end{equation}
for Matern72, the covariance function is
\begin{align}\label{equ:GP_covf_Matern72}
\nonumber k(x,\tilde{x}) &=\sigma_f^2 \exp\left[-\frac{\sqrt{7}|x - \tilde{x}|}{\ell} \right] \left(1+\frac{\sqrt{7}|x - \tilde{x}|}{\ell} \right.\\ 
& \left. +\frac{14(x - \tilde{x})^2}{5\ell^2} 
+\frac{7\sqrt{7}|x - \tilde{x}|^3}{15\ell^3} \right),
\end{align}
and for Matern92, the covariance function is
\begin{align}\label{equ:GP_covf_Matern92}
\nonumber k(x,\tilde{x}) &=
\sigma_f^2 \exp\left[-\frac{3|x - \tilde{x}|}{\ell} \right] \left(1+\frac{3|x - \tilde{x}|}{\ell} +\frac{27(x - \tilde{x})^2}{7\ell^2} \right.\\
&\left. +\frac{18|x - \tilde{x}|^3}{7\ell^3} +\frac{27(x - \tilde{x})^4}{35\ell^4} \right),
\end{align}
where $\sigma_f$ and $\ell$ are two hyperparameters that should be optimized. With the same procedure of sections \ref{subsec:rec_with_GP} and \ref{sec:omk_from_Hz_SNe_gp}, we reconstruct functions of $H(z)$ by adopting these four kinds of covariance functions and constrain the cosmic curvature $\Omega_K$ by comparing $\mu_H$ and the distance modulus obtained from the Pantheon SNe Ia. $1\sigma$ constraints on $\Omega_K$ are shown in Table \ref{tab:omk_gp_multiCov}, and the corresponding one-dimensional distribution of $\Omega_K$ are shown in Figure \ref{fig:omk_gp_multiCov}. 

\begin{table*}
	\centering\caption{$1\sigma$ Constraints on $\Omega_K$ constrained from $H(z)$ and Pantheon SNe Ia. The $H(z)$ is reconstructed by using the GP method with the Matern class of covariance functions. Case (a), (b) and (c) represent no prior of $H_0$, $H_0 = 67.74\pm 0.46~\rm km~ s^{-1} ~Mpc^{-1}$, and $H_0 = 73.24\pm 1.74~\rm km~ s^{-1} ~Mpc^{-1}$ when reconstructing the function of $H(z)$, respectively.}\label{tab:omk_gp_multiCov}
	\begin{tabular}{c|c|c|c}
		\hline\hline
		Covariance functions & Case (a) & Case (b) & Case (c) \\
		\hline
		Matern32 & $0.203\pm0.203$ & $0.305\pm0.205$ & $-0.445\pm0.233$ \\
		Matern52 & $0.210\pm0.193$ & $0.342\pm0.220$ & $-0.444\pm0.235$ \\
		Matern72 & $0.239\pm0.197$ & $0.340\pm0.208$ & $-0.454\pm0.243$ \\
		Matern92 & $0.275\pm0.202$ & $0.333\pm0.214$ & $-0.445\pm0.233$ \\
		\hline\hline
	\end{tabular}
\end{table*}

Obviously, for all the four kinds of covariance functions, we can see that the results of case (a) are similar to that of case (b), and both of them are greatly different from that of case (c). This means that the constraint on $\Omega_K$ is greatly influenced by the prior of $H_0$, which is similar to the result of section \ref{sec:omk_from_Hz_SNe_gp} (see the right panel of Figure \ref{fig:omk_ann_gp}). Despite the covariance function has a little influence on final constraint on the cosmic curvature, the effect of the prior of $H_0$ on the result seems an insurmountable problem for the GP method. Therefore, the GP method should be used with caution in the reconstruction of $H(z)$.

\section{Conclusions}\label{sec:conclusion}

In this work, we reconstruct functions of CC $H(z)$ with the ANN and GP methods and integrate them to obtain the distance modulus $\mu_H$. Then we constrain the cosmic curvature $\Omega_K$ by comparing $\mu_H$ and the distance modulus obtained from Pantheon SNe Ia. We find that the function of $H(z)$ reconstructed by GP can be greatly influenced by the prior of the Hubble constant. However, the ANN method can overcome this to reduce the influence of the prior of the Hubble constant on the function of $H(z)$, and further reduce the influence on the measurement of the cosmic curvature. Therefore, the ANN method may surpass the GP method in the measurement of the cosmic curvature. 

Based on the ANN method, we find a positive value of $\Omega_K$ is favored by the current CC $H(z)$ and Pantheon SNe Ia data, and the difference between this result and that obtained using Planck CMB is $1.6\sigma$. In order to test the reliability of the ANN method in the measurement of the cosmic curvature, we further constrain the cosmic curvature in a model-independent way, by using the simulated Hubble parameter and the GW standard sirens that from the Einstein Telescope. We find the ANN method is reliable and unbiased, and thus the deviation of the cosmic curvature from the flat universe is no caused by the ANN method.

Moreover, the results show that the error of $\Omega_K$ is $\sim0.186$ when 100 GW events with electromagnetic counterparts are detected, which is $\sim56\%$ smaller than that constrained from the $H(z)$ and Pantheon SNe Ia, and $\sim0.056$ when having 1000 GW events with electromagnetic counterparts. Therefore, the data-driven method based on ANN has potential in the measurement of the cosmic curvature when using the future Hubble parameter and GW standard siren.

\section{Acknowledgement}
We thank Zhengxiang Li and Jingzhao Qi for helpful discussions. This work is supported by the National Science Foundation of China under grants No. U1931202 and 12021003, and the National Key R\&D Program of China under grant No. 2017YFA0402600.

\section{Data availability}

The $H(z)$ data using in this paper are available in \citet{Wanggj:2020}, the Pantheon SNe Ia data are available in \citet{Scolnic:2018} and in its online supplementary material (doi:\href{https://doi.org/10.17909/T95Q4X}{10.17909/T95Q4X}), and the simulated GW data will be shared on reasonable request to the corresponding author.

\bsp
\label{lastpage}
\end{document}